\documentclass[aps,10pt,twocolumn,pra,showpacs,amsmath,amsfonts,amssymb,longbibliography]{revtex4-1} 
\usepackage{graphicx}
\usepackage{verbatim}
\usepackage{bbm} 
\usepackage{xcolor}
\newcommand{\beq}{\begin{equation}} 
\newcommand{\eeq}{\end{equation}}                   
\newcommand{\ra}{\rangle}
\newcommand{\la}{\langle}

\newcommand{\tr}{{\rm Tr}}
\begin{document}

\title{Quantum information theory of the Bell-state quantum eraser}

\author{Jennifer R. Glick}
\email[]{patte399@msu.edu}

\author{Christoph Adami}
\email[]{adami@msu.edu}

\affiliation{Department of Physics and Astronomy, Michigan State University, East Lansing, Michigan 48824, USA}

\date{\today} 

\begin{abstract}
Quantum systems can display particle- or wave-like properties, depending on the type of measurement that is performed on them. The Bell-state quantum eraser is an experiment that brings the duality to the forefront, as a single measurement can retroactively be made to measure particle-like or wave-like properties (or anything in between). Here we develop a unitary information-theoretic description of this (and several related) quantum measurement situations that sheds light on the trade-off between the quantum and classical features of the measurement. In particular, we show that both the coherence of the quantum state and the classical information obtained from it can be described using quantum-information-theoretic tools only, and that those two measures satisfy an equality on account of the chain rule for entropies. The coherence information and the which-path information have simple interpretations in terms of state preparation and state determination, and suggest ways to account for the relationship between the classical and the quantum world.  

\end{abstract}

\maketitle

\section{Introduction}
Wave-particle duality is an iconic feature of quantum mechanics, one not shared by classical systems in which an object cannot simultaneously have a wave and particle nature. Unraveling the mysteries behind wave-particle duality has occupied the better part of the last century, while significant advances in our understanding have come both from clever experimental approaches as well as theoretical developments. Pivotal experiments were the quantum optical implementations of Wheeler's~\cite{Wheeler1978} delayed-choice experiment~\cite{Jacquesetal2007,Peruzzoetal2012,kaiser2012} (which are equivalent in principle to delayed-choice quantum eraser experiments~\cite{herzog1995,Kimetal2000}), while the theoretical advances have framed the discussion of the wave-particle duality in terms of a quantum-information-theoretic trade-off between the coherence of the quantum system (or ``quanton") and the information that one may attempt to obtain about the particle path in the interferomater or double-slit experiment (the ``which-path" information)~\cite{greenberger1988,englert1996,Colesetal2014,AngeloRibeiro2015,Baganetal2016}.

The delayed-choice experiments highlight an important feature of this new understanding of wave-particle duality: while as per Bohr's complementarity principle~\cite{Bohr1984} it is the nature of the experiment that determines whether we shall observe wave- or particle-like behavior, it is clear that the nature of the experiment can be changed after it has already taken place. In other words, the same experiment can retroactively be made to measure wave or particle properties, or anything in between~\cite{AharonovZubairy2005,Peruzzoetal2012,kaiser2012}. Such a state of affairs is often greeted with skepticism, as the experiments seem (to some) to imply that the delayed choice of the measurement changes the quantum state retroactively, thus violating causality (see, e.g., the discussion in~\cite{AharonovZubairy2005}). The natural interpretation of these results is that a quanton has both particle-like and wave-like properties at the same time, and that the results of measurements can reveal one or a mixture of those characteristics. Due to the classical nature of the measuring devices, however, they do not---in fact, cannot---reflect the true nature of the quantum state. In the following, we will make these arguments in a strictly information-theoretic setting. 

We develop the framework of (possibly delayed) complementary measurements (which-path or which-phase) in terms of a quantum information-theoretic description of the Bell-state quantum eraser, but the formalism is general and applies equally to any situation where measurements are made in parallel on two (and in an obvious extension to several) entangled quantum systems, such as the Garisto-Hardy entanglement eraser~\cite{GaristoHardy1999}. 

We first describe the ordinary double-slit experiment performed on one half of a Bell state, then the polarization-tagged version where which-path information can be extracted, followed by the quantum erasure procedure. In the next section we describe these steps in terms of classical and quantum information theory that results in an information-theoretic equality that mirrors (and is completely analogous to) the trade-off between distinguishability and visibility of Greenberger and Yasin~\cite{greenberger1988} as well as Englert~\cite{englert1996}. The equality involves the coherence of the quanton and the information obtained about its path just as Bagan et al.\ have recently shown~\cite{Baganetal2016}, but we do not assume a specific form for the measure of coherence as it emerges naturally from the information-theoretic analysis. The ``conservation law" between coherence and information appears simply as a consequence of the chain rule for entropies. We offer conclusions in which we suggest what information is actually encoded in the measurement devices, given that it cannot reflect information about the quantum state.

\section{The Bell-State Quantum Eraser}
\begin{figure}[h]
  \centering
  \includegraphics[width=0.925\linewidth]{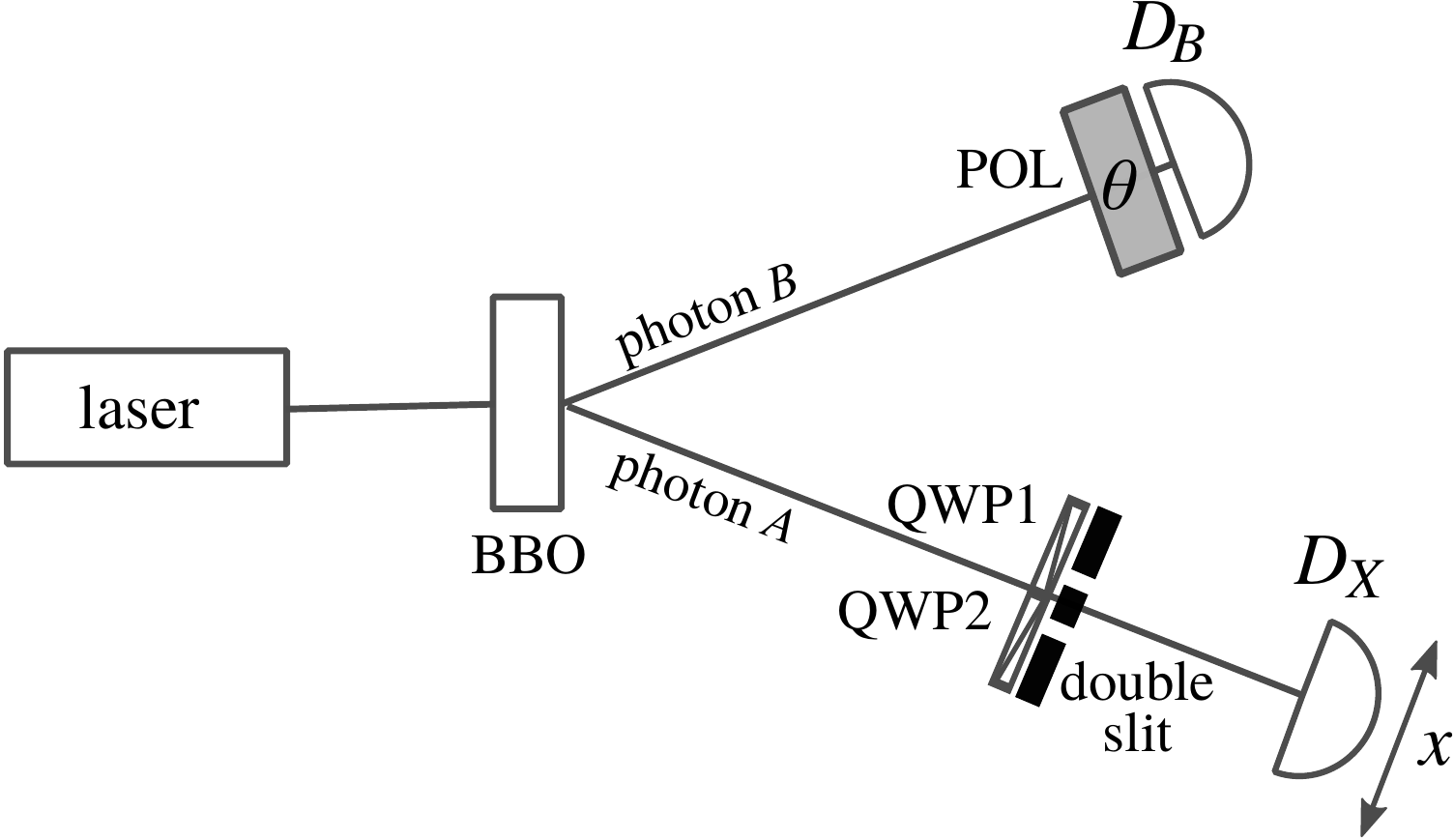} 
  \caption{Schematic of the double-slit Bell-state quantum eraser experiment~\cite{walborn2002}. After production of the Bell state~\eqref{eraser-bell} by type-II spontaneous parametric down-conversion (SPDC) in the BBO ($\beta$-barium borate) crystal, photon $B$ travels along the upper branch to a linear polarizer (POL, set to angle $\theta$ relative to the $|h\ra,\, |v\ra$ basis) and detector $D_B$. Photon $A$ travels down to the quarter-wave plates (QWPs) and double slit, and then to detector $D_X$, which plays the role of a ``screen'' by scanning along the direction $x$.}
  \label{fig:schematic}
\end{figure}

The Bell-state quantum eraser experiment~\cite{walborn2002}, as illustrated in Fig.~\ref{fig:schematic}, proceeds as follows. By spontaneous parametric down-conversion (SPDC)~\cite{herzog1995}, a pair of photons $A$ and $B$ are prepared that are entangled in a Bell state~\footnote{Note
that an obvious extension of the present construction is to allow for arbitrary entangled photon pairs, such as for example the q-Bell states~\cite{AdamiCerf1997} in which a parameter $q$ interpolates between product states ($q$=0 or 1) and fully entangled states ($q=1/2$). In that case, certain entries that are 1 bit in the following would be replaced by $H[q]=-q\log q -(1-q)\log(1-q)$.}

\beq \label{eraser-bell}
    |\Psi\ra_{AB} = \frac{|h\ra |v\ra + |v\ra |h\ra}{\sqrt2}~, 
\eeq
where the first and second states refer to photons $A$ and $B$, respectively, and $|h\ra,|v\ra$ are the horizontal and vertical linear polarization states. Photon $B$, called the ``idler'', is sent in one direction towards a photodetector $D_B$. A polarization filter (denoted POL in Fig.~\ref{fig:schematic}) can be placed in its path to perform measurements in specific bases. Meanwhile, photon $A$, called the ``signal'', is sent in another direction towards a double-slit apparatus. Photon $A$ will pass through the double slit and subsequently be detected by a photodetector $D_X$ that scans along the $x$-axis. From the detector counts of $D_X$ one can construct an interference pattern as a function of the spatial variable $x$. The interference pattern is erased by placing two quarter-wave plates (QWPs) in front of each slit. This tags the path of photon $A$ and provides path information. See Fig.~\ref{fig:schematic} for a schematic of the experiment.

\subsection{Splitting the photon}
 The full wavefunction describing the entangled pair of photons $A$ and $B$ is
 \beq\label{bellstate}
    |\Psi\ra_{AB} = \frac{|h\ra_P |v\ra_B + |v\ra_P |h\ra_B}{\sqrt2}  \otimes |\psi\ra_Q ,
 \eeq 	
 where the Hilbert space $\mathcal{H}_A = \mathcal{H}_P \otimes \mathcal{H}_Q$ of photon $A$ is composed of polarization $P$ and spatial $Q$ degrees of freedom ($Q$ is called the quanton). The polarizations of photons $A$ and $B$, entangled in a Bell state, are decoupled from the spatial state $|\psi\ra$ of photon $A$. We drop the spatial part of photon $B$ as it remains decoupled throughout. Sending photon $A$ through the double slit transforms only the spatial states of $A$ so that Eq.~\eqref{bellstate} evolves to
 \beq
    |\Psi\ra_{AB} = \frac{|h\ra_P |v\ra_B + |v\ra_P |h\ra_B}{\sqrt2}  \otimes \frac{|\psi_1\ra_Q+|\psi_2\ra_Q}{\sqrt2}.
 \eeq 	 
 The states $|\psi_j\ra$ denote the path of photon $A$ corresponding to slit $j$~\footnote{Note that the extension of this framework to allow for $N$-path devices is straightforward.}.  The spatial degree of freedom of $A$, denoted by $Q$, is still independent from its polarization. 

 Tracing over the states of photon $B$, the density matrix describing the spatial modes of photon $A$ is the pure state
 \beq
    \rho_{Q} \!=\! \frac12 \Big( |\psi_1\ra_Q\la \psi_1| +|\psi_2\ra_Q\la\psi_2| + |\psi_1\ra_Q\la \psi_2| + |\psi_2\ra_Q\la \psi_1| \Big).
 \eeq 
 The expectation value in the position basis $|x\ra$ of the screen $D_X$ yields the probability to observe photon $A$ at position $x$
 \beq
    \la x | \rho_{Q} |x \ra = p(x) = \frac12 \big| \psi_1(x)+\psi_2(x)\big|^2,
 \eeq 
 where we define the amplitudes $\psi_j(x) = \la x | \psi_j\ra$. This probability distribution is a coherent superposition and the usual double-slit interference fringes will be observed on the screen. In the appendix we show how the characteristic fringes can be derived from a von Neumann measurement of $Q$ by the detector $D_X$.

 \subsection{Tagging the photon path}
 To extend this discussion to a quantum eraser experiment, a tagging operation is performed on the two branches of the double-slit apparatus in order to provide information about the path of photon $A$. In practice, this is implemented by placing a quarter-wave plate (QWP) in front of each slit. The Jones matrix
 for a general wave plate (WP) oriented at an angle $\beta$ (the fast axis) to our coordinate system (in this case, $|h\ra$ and $|v\ra$) is~\cite{jones1941,collett2005}
 \beq 
    U \!\!=\!\! \begin{pmatrix}
           \!\cos (\frac\alpha2) \!+\! i \sin (\frac\alpha2) \cos (2 \beta) & i \sin (\frac\alpha2) \sin (2 \beta) \\
           i \sin (\frac\alpha2) \sin (2 \beta) & \cos (\frac\alpha2) \!-\! i \sin (\frac\alpha2) \cos (2 \beta)\!
         \end{pmatrix}\!,
 \eeq 
 where $\alpha = \pi/2$ for a QWP. More specifically, the QWP in front of slit 1 (slit 2) has its fast axis at $\beta = 45$ ($\beta = -45$)
 degrees, leading to
 \beq 
    U^{(\pm)}_{QWP} = \frac{1}{\sqrt2} \begin{pmatrix}
           1 & \pm i \\
           \pm i & 1
         \end{pmatrix},
 \eeq
 where $U^{(+)}_{QWP} \!=\! U^{(1)}_{QWP}$ and $U^{(-)}_{QWP} \!=\! U^{(2)}_{QWP}$ are the matrices associated with slit 1 and 2, respectively. These transform linearly polarized states $|h\ra$ and $|v\ra$ into circularly polarized states according to
 \begin{eqnarray} 
    & &U^{(1)}_{QWP} ~ |h\ra = \frac{|h\ra + i |v\ra}{\sqrt2} = |L\ra, \nonumber\\
    & &U^{(1)}_{QWP} ~ |v\ra = \frac{|v\ra + i |h\ra}{\sqrt2} = i |R\ra, \nonumber\\
    & &U^{(2)}_{QWP} ~ |h\ra = \frac{|h\ra - i |v\ra}{\sqrt2} = |R\ra,\nonumber\\
    & &U^{(2)}_{QWP} ~ |v\ra = \frac{|v\ra - i |h\ra}{\sqrt2} = -i |L\ra,\nonumber
 \end{eqnarray} 	
 where $|R\ra$ ($|L\ra$) denotes right-handed (left-handed) circular polarization.

\begin{figure}[h]
    \includegraphics[width=0.895\linewidth]{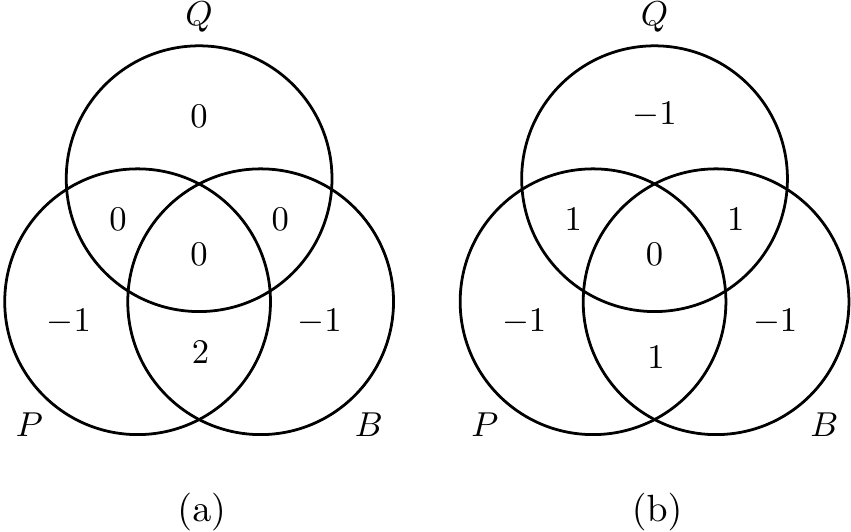}
    \caption{Entropic Venn diagrams showing the effect of the tagging operation. (a) Before tagging, the polarization states of $A$ and $B$ are entangled in a Bell state. (b) After tagging, the spatial degree of freedom of photon $A$ becomes entangled with the polarization of $A$ and $B$ according to Eq.~\eqref{tagged}.}
    \label{fig:venn-set1}
 \end{figure}

 When photon $A$ passes through the QWPs and the double slit, its polarization becomes entangled with its spatial degree of freedom so that the wavefunction~\eqref{bellstate} evolves to 
 \beq
 \begin{split}
   |\widetilde{\Psi}\ra_{\!AB} = \frac{1}{\sqrt2}  \Bigg( &  \frac{|L\ra_{P} |v\ra_{B} + i \, |R\ra_{P} |h\ra_{B}}{\sqrt2} \otimes |\psi_1\ra_Q \\
      & +  \frac{|R\ra_{P} |v\ra_{B} - i \, |L\ra_{P} |h\ra_{B}}{\sqrt2} \otimes |\psi_2\ra_Q \! \Bigg) ,
 \end{split}
 \eeq 	 
 where the tilde indicates that the tagging operation has been performed. Grouping together the polarization states of photon $B$, we can equivalently express this state as
 \beq\label{tagged}
 \begin{split}
   |\widetilde{\Psi}\ra_{\!AB} = \frac{1}{\sqrt2} \Bigg( & \frac{ \!|\psi_1\ra_Q |L\ra_{P} + |\psi_2\ra_Q |R\ra_{P} }{\sqrt2} \otimes |v\ra_{B} \\
      & \! + i \, \frac{|\psi_1\ra_Q |R\ra_{P}  -  |\psi_2\ra_Q |L\ra_{\!P}}{\sqrt2} \otimes |h\ra_{B} \! \Bigg) .
 \end{split}
 \eeq 
 The entanglement between the two degrees of freedom of photon $A$ causes the spatial modes $Q$ to become completely mixed
 \beq\label{rhoAs} 
    \rho_Q = \frac12 \big(|\psi_1\ra_Q\la\psi_1| + |\psi_2\ra_Q\la\psi_2|\big), 
 \eeq 
 so that interference is no longer observed on the screen. In Fig.~\eqref{fig:venn-set1} we show the entropic Venn diagrams~\cite{CerfAdami1998} before (a) and after (b) the tagging operation with the QWPs. In these diagrams, the sum of all the entries in a circle add up to the entropy of the subsystem, and the entropy shared between subsystems is indicated in the overlap between circles. Conditional entropies appear in unshared areas of the circle, and can be negative in quantum mechanics~\cite{CerfAdami1997} (they must be positive if they are classical Shannon entropies). Entropies shared between three systems (the center of the diagrams in Fig.~\ref{fig:venn-set1}) can be negative both in classical and quantum physics~\cite{CerfAdami1998}. All of the von Neumann entropies $S(\rho)=-\tr \rho\log \rho$ can be calculated in a straightforward manner from the density matrix $\rho_{QPB}=|\widetilde{\Psi}\ra_{\!AB}\la \widetilde{\Psi}|$ and the marginalized density matrices $\rho_Q=\tr_{PB}(\rho_{QPB})$, $\rho_B=\tr_{QP}(\rho_{QPB})$, etc.
 
 Before tagging (see Eq.~\eqref{bellstate}), $Q$ is completely decoupled from the polarization $P$ of photon $A$ and of photon $B$, which together are entangled in a Bell state. After tagging (see Eq.~\eqref{tagged}), all three variables $Q$, $P$, and $B$ are in a tripartite entangled state. Note that the ternary mutual entropy $S(Q:P:B)$ vanishes in both diagrams since the underlying density matrix is a pure state~\cite{CerfAdami1998}. The expression of the ternary shared entropy in terms of sub-system entropies can be read off the Venn diagram in general as $ S(Q:P:B)=S(Q)+S(B)+S(B)-S(QB)-S(QP)-S(PB)+S(QPB)$, and similarly for any pair-wise shared entropies.

 \subsection{Erasing the photon path}
 As is by now well-known~\cite{Scullyetal1991}, it is still possible to extract an interference pattern from the screen data, even when the quanton has been tagged, {\em}if we have additional information about the state of photon $B$. Suppose we perform a measurement of $B$ in a basis that is described by an angle $\theta$ relative to the $|h\ra,\,|v\ra$ basis. For a general change of the basis, the polarization states of $B$ are written as
 \begin{eqnarray}
      &|v\ra_B = U_{00} |0\ra_B + U_{01} |1\ra_B, \\
      &|h\ra_B = U_{10} |0\ra_B + U_{11} |1\ra_B.
 \end{eqnarray}
 For simplicity, we use the following parameterization for the matrix elements $U_{ij}$ of the rotation operator that transforms $|h\ra,|v\ra$ to the new basis spanned by $|0\ra,|1\ra$ in terms of the single angle $\theta$:
 \beq \label{U}
    U = \begin{pmatrix}
          \cos \theta & - \sin\theta \\
          \sin \theta & ~\,\,\cos\theta
        \end{pmatrix}.
 \eeq 
 Rewriting the states of $B$ in the basis $|0\ra,|1\ra$ and entangling $B$ with a polarization detector $D_B$ transforms Eq.~\eqref{tagged} to~\cite{CerfAdami1998}
 \beq\label{AB-DB}
        |\widetilde{\Psi}\ra_{\!ABD_{B}\!\!} = \frac12 \sum_{mk} i^m \,\, |\psi^k_m\ra_Q \otimes |m\ra_{P} \otimes |kk\ra_{BD_{B}},
 \eeq
 where $k=0,1$ labels the polarization of $B$ and $D_B$ while $m=0$\,(``$L$''),\,1\,(``$R$'') denotes the polarization of $A$,
 and where we defined the spatial states $|\psi_m^k\ra_Q$ of photon $A$ 
 \begin{eqnarray}
      & |\psi_L^k\ra_Q = U_{0k} |\psi_1\ra_Q - i U_{1k} |\psi_2\ra_Q, \label{psi-kL}\\
      & |\psi_R^k\ra_Q = U_{1k} |\psi_1\ra_Q - i U_{0k} |\psi_2\ra_Q. \label{psi-kR}
 \end{eqnarray}
 These states thus describe the spatial state of photon $A$ (the quanton) when it has a circular polarization $m$ and is correlated with photon $B$ that has polarization $k$. 
 Only the states for a given polarization $m$ are orthonormal
 \begin{eqnarray}
      \la \psi^{k'}_m|\psi^{k}_m\ra & = & \delta_{kk'}, \\
      \la \psi^{k'}_L|\psi^{k}_R\ra & = & U_{0k'}^* \, U_{1k} + U_{0k} \, U_{1k'}^* .
 \end{eqnarray}
 Of course, the quanton state $\rho_Q$ derived from~\eqref{AB-DB} is still completely mixed as in~\eqref{rhoAs} so that no interference can be observed on the screen. However, as long as the erasure angle $\theta$ is nonzero it is now possible to extract an interference pattern given the outcome of the polarization measurement of photon $B$ (even if that measurement occurs much later than the measurement of photon $A$).

 From the wavefunction~\eqref{AB-DB}, we can compute the joint density matrix for photon $A\equiv QP$ (spatial $Q$ and polarization $P$) and detector $D_B$. Tracing over $B$ yields
 \beq\label{rhoAsApDB}
 \begin{split}
    \rho_{AD_{B}\!} 
             & = \!\frac12 \sum_k \rho_A^k \otimes |k\ra_{\!D_{B}\!}\la k|,
 \end{split}
 \eeq 
 where the (orthonormal) states of photon $A$, conditional on the state $k$ of detector $D_B$, are $\rho_A^k = |\phi_k\ra\la\phi_k|$ and $|\phi_k\ra = \frac{1}{\sqrt2} \sum_m i^m \, |\psi^k_m\ra_Q \otimes |m\ra_P$. The effect of the erasure is contained in the behavior of these states as the measurement angle $\theta$ is varied. For a measurement in the original basis ($\theta = 0$), detector $D_B$ prepares photon $A$ in one of the fully entangled states: $|\phi_0\ra \propto |\psi_1\ra_Q |L\ra_{P} + |\psi_2\ra_Q |R\ra_{P\!}$ and $|\phi_1\ra \propto |\psi_1\ra_Q |R\ra_{P} - |\psi_2\ra_Q |L\ra_{P\!}$. From these expressions we can infer, with a polarization measurement of photon $A$, its path. For instance, outcome $k=0,\,m=L$ is associated with the spatial state $|\psi_1\ra$ for slit 1. Therefore, polarization measurements of photon $B$ at $\theta=0$ yield full path information and no interference fringes.
 
 On the other hand, for a measurement in the diagonal basis at $\theta = \pi/4$, detector $D_B$ prepares photon $A$ in one of the completely decoupled states: $|\phi_0\ra \propto (|\psi_1\ra_Q-i |\psi_2\ra_Q) \otimes (|L\ra_{P\!} + i |R\ra_{P})$ and $|\phi_1\ra \propto (|\psi_1\ra_Q + i |\psi_2\ra_Q) \otimes (-|L\ra_{P\!} + i |R\ra_{P})$. Now, a polarization measurement of $A$ cannot reveal path information, and the coherently summed spatial modes lead to interference fringes. In the appendix, we compute these interference patterns and show their dependence on the erasure angle $\theta$. Thus, we see that, regardless of the temporal order of the two polarization measurements, the measurement of $B$ can be seen as state {\em preparation}, while the measurement of $A$ is state {\em determination}, that is, extraction of which-path information.

 \section{Information theory}
 \subsection{State preparation}
 We illustrate the quantum erasure mechanism by building on the entropic Venn diagrams in Fig.~\ref{fig:venn-set1} and constructing the entropic relationships between the random variables $Q$, $P$, and $D_B$ from the joint and marginal entropies associated with Eq.~\eqref{rhoAsApDB}. In the basis $|\phi_k\ra \otimes |k\ra$ it is clear that the entropy of~\eqref{rhoAsApDB} is $S(QPD_B) = 1$, while the marginal entropies are all $S(Q) = S(P) = S(D_B) = 1$. Tracing over detector $D_B$, the total entropy of $A\equiv QP$ is also $S(QP) = 1$. The pairwise entropy $S(PD_B)$ is equal to $S(QB)$ by the Schmidt decomposition of~\eqref{AB-DB}, which in turn is equal to $S(QD_B)$ by the symmetry between the $B$ and $D_B$ states. The remaining pairwise entropy is then computed from the density matrix
 \beq\label{rhoAsDB}
 \begin{split}
    \rho_{QD_B} 
       & = \frac12 \sum_k \rho^k_Q \otimes |k\ra_{\!D_B\!}\la k|,
 \end{split}
 \eeq 
 where the state of $Q$ conditional on the outcome $k$ of detector $D_B$ is
 \beq\label{rhoAs-k}
 \begin{split} 
    \rho^k_Q & = \frac12 \sum_{m} |\psi^k_m\ra_Q\la\psi^k_m| \\
       & = \frac12 \Big( |\psi_1\ra_Q\la\psi_1| +  |\psi_2\ra_Q\la\psi_2| \\
       & \qquad~~ + i \,(-)^k \sin2\theta \Big[ |\psi_1\ra_Q\la\psi_2| -  |\psi_2\ra_Q\la\psi_1| \Big] \Big).
 \end{split}
 \eeq 
 From this expression, we see that for a given outcome $k$, the spatial degree of freedom of photon $A$ is generally no longer mixed (as in Eq.~\eqref{rhoAs}) and has a coherence that is controlled by the sine of the measurement angle. In turn, it is possible to extract interference fringes from the screen. With the Bloch vector $\vec{a} \!=\! -(-)^k \sin(2\theta) \, \hat{y}$, Eq.~\eqref{rhoAs-k} can be expressed as $\rho^k_Q = \frac12 ( \mathbbm{1} - (-)^k \sin(2\theta) \, \hat{\sigma}_y )$, where $\hat{y}$ is a unit vector and $\sigma_y$ is a Pauli matrix. This density matrix varies from a fully mixed state $|\vec{a}| = 0$ at $\theta=0$ to a pure state $|\vec{a}| = 1$ at $\theta = \pi/4$. 
 
 To compute the entropy of the block-diagonal matrix in Eq.~\eqref{rhoAsDB}, we first find the entropy of $\rho^k_Q$. The eigenvalues of $\rho^k_Q$ are $\lambda_{\pm} \!=\! \frac12(1 \pm |\vec{a}|) \!=\! \frac12( 1 \pm \sin (2\theta))$ and are independent of the index $k$, leading to equal entropies $S(\rho^0_Q) = S(\rho^1_Q)$. Therefore, the entropy of~\eqref{rhoAsDB}, which is also equal to the entropy $S(PD_B)$, is~\cite{NielsenChuang_Book}
 \beq \label{sqdb}
 \begin{split}
    S(QD_B) & = 1 + \frac12 \sum_k S(\rho^k_Q) = 1+S, \\
 \end{split}
 \eeq 
 where $S = -\lambda_+ \log\lambda_+  - \lambda_- \log\lambda_-$ is the entropy of~\eqref{rhoAs-k}, and varies from $S = 1$ at $\theta = 0$ to $S = 0$ at $\theta = \pi/4$.

 The relationship between the variables $Q$, $P$, and $D_B$ is summarized by the entropic Venn diagram in Fig.~\ref{fig:venn}(a). As a result of tagging the path of photon $A$, the spatial and polarization modes of $A$ are entangled, given the state of $D_B$, for $S > 0$ ($\theta < \pi/4$). The amount of entanglement $S$
 varies with the erasure angle $\theta$ and specifies the degree to which the polarization $P$ can reveal information about the spatial mode $Q$. The non-zero ternary mutual entropy $S(Q\!:\!P\!:\!D_B) = 1-2S$ indicates that the mutual entropy of $Q$ and $P$ can be shared by detector $D_B$~\cite{AdamiCerf1997}.

 \subsection{State determination}
 In order to reveal information about the path of photon $A$, its polarization is measured after it passes the double-slit apparatus using a detector $D_A$ in the circular $|L\ra,|R\ra$ basis. After the measurement with $D_A$, the wavefunction~\eqref{AB-DB} evolves to
 \beq\label{AB-DADB}
    |\Psi'\ra_{\!AD_{A}BD_{B}\!\!} =\! \frac12 \!\sum_{mk} i^m  |\psi^k_m\ra_{Q} \otimes |mm\ra_{\!PD_{A}} \!\otimes |kk\ra_{\!BD_{B}}.\!
 \eeq 
 The entropic relations between the variables $Q$, $D_A$, and $D_B$ are computed from their joint density matrix, which is found by tracing out the polarization states of photons $A$ and $B$ from~\eqref{AB-DADB}, 
 \beq\label{As-DADB}
    \rho_{QD_AD_B} \!=\! \frac14 \sum_{mk} |\psi^k_m\ra_Q\la\psi^k_m| \otimes |m\ra_{\!D_A\!}\la m| \otimes |k\ra_{\!D_B\!}\la k|.
 \eeq 
 For a set of outcomes $k$ and $m$, the corresponding spatial state of photon $A$ is $|\psi^k_m\ra_Q$. It is straightforward to show that the entropy of Eq.~\eqref{As-DADB} is $S(QD_A D_B) = 2$, while the marginal entropies are $S(Q) = S(D_A) = S(D_B) = 1$.
 
 Tracing over the spatial states $Q$ in Eq.~\eqref{As-DADB}, we find that the two polarization detectors are uncorrelated 
 \beq 
    \rho_{D_A D_B} = \frac12 \mathbbm{1}_{D_A} \otimes \frac12 \mathbbm{1}_{D_B},
 \eeq 
 with a joint entropy $S(D_A D_B) = 2$ and  
 \beq
    S(D_A:D_B) = 0.
 \eeq 
 Thus, the measurement with $D_A$ reveals no information about the state of $D_B$. This is not surprising as the QWPs act as a ``controlled-NOT" gate on the polarization. 
 Indeed, conditioning on the spatial states of photon $A$ yields
 \beq
    S(D_A:D_B|Q) = S \ge 0. 
 \eeq  
 In a sense, therefore, the quanton states encrypt the relationship between the state preparation with $D_B$ and its readout with $D_A$.

 Tracing out the states of the polarization detector $D_A$ in Eq.~\eqref{As-DADB}, the joint density matrix $\rho_{Q D_B}$ is unchanged from Eq.~\eqref{rhoAsDB}, since the measurement with $D_A$ does not affect the correlations between $Q$ and $D_B$. 

\begin{figure}[t]
      \centering
      \includegraphics[width=0.95\linewidth]{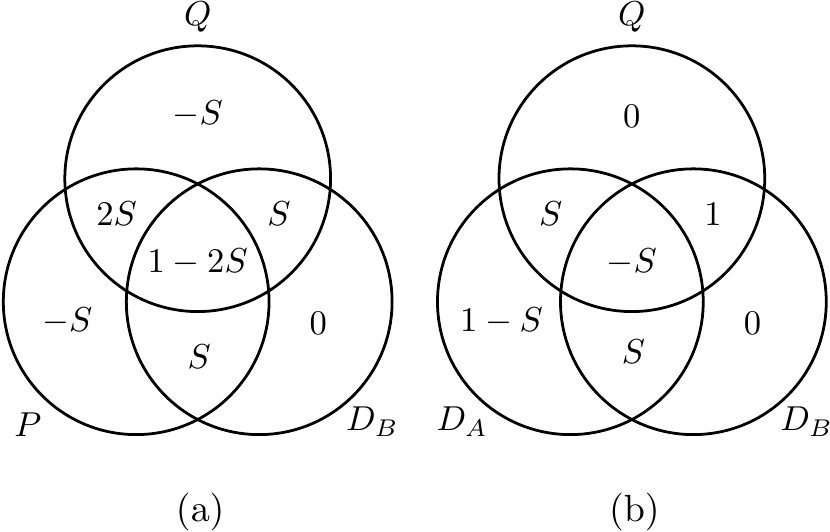}
      \caption{
      Entropic Venn diagrams for (a) state preparation with detector $D_B$ (see Eq.~\eqref{AB-DB}), and (b) state determination with detector $D_A$ (see Eq.~\eqref{AB-DADB}).
      \label{fig:venn}}
 \end{figure}

 Tracing over the states of $D_B$ in Eq.~\eqref{As-DADB}, the joint density matrix for $Q$ and $D_A$ in turn is
 \beq\label{rhoAsDA}
 \begin{split}
    \rho_{Q D_A} 
       & = \frac12 \sum_{m} \rho^m_Q\otimes |m\ra_{\!D_A\!}\la m| = \frac12 \mathbbm{1}_Q \otimes \frac12 \mathbbm{1}_{D_A},
 \end{split} 
 \eeq  
 where the density matrix of $Q$, conditional on the polarization outcome $m$ of detector $D_A$, is 
 \beq
    \rho^m_Q \!=\! \frac12 \!\sum_k |\psi^k_m\ra_Q\la\psi^k_m| \!=\! \frac12 \Big( |\psi_1\ra_Q\la\psi_1| +  |\psi_2\ra_Q\la\psi_2|\Big),\!
 \eeq 
 which is independent of the polarization index $m$, and is equivalent to the full density matrix $\rho_Q$. That is, $\rho^m_Q = \rho_Q = \frac12 \mathbbm{1}_Q$, and is a completely mixed state. Finally, the joint entropy of $Q$ and $D_A$ is $S(Q D_A) = 2$. The entropic relationships between the variables $Q$, $D_A$ and $D_B$ are summarized by the Venn diagram in Fig.~\ref{fig:venn}(b). 
 
 
 \subsection{Information-theoretic origins of coherence and path information}
 From the marginal and joint entropies computed in the previous section, we can construct information-theoretic relationships between the variables $Q$, $D_A$, and $D_B$.  First, as previously noted, detectors $D_B$ and $D_A$ share no information 


 {\em Coherence.---}The information shared between the preparation with detector $D_B$ and the quanton $Q$ (the spatial state of photon $A$) is given by the mutual entropy
 \beq
    S(Q : D_B) = 1 - S \le 1\;,
 \eeq 
 with $S$ defined in Eq.~(\ref{sqdb}).  
 We can understand how this entropy depends on the erasure angle $\theta$ by considering two cases. First, from the joint density matrix $\rho_{Q D_B}$ in Eq.~\eqref{rhoAsDB}, a measurement of photon $B$ at an angle $\theta = 0$ decouples $Q$ from detector $D_B$
 \beq
    \theta = 0\!: ~~\rho_{Q D_B} = \frac12 \mathbbm{1}_Q \otimes \frac12 \mathbbm{1}_{D_B},
 \eeq
 and the conditional state $\rho^k_Q$ in Eq.~\eqref{rhoAs-k} becomes a statistical mixture, i.e., interference cannot be observed on the screen. In this case, $S(Q : D_B) = 0$ and there is no information shared between the two variables. However, increasing the erasure angle to $\theta = \pi/4$ leads to perfect correlation
 \beq
    \theta \!=\! \frac{\pi}{4}\!: ~~\rho_{Q D_{\!B\!\!}} = \!\frac12 \Big(|f\ra_Q\la f| \otimes |0\ra_{\!D_{\!B}\!}\la 0| + |a\ra_Q\la a|\otimes |1\ra_{\!D_{\!B}\!}\la1| \Big), 
 \eeq
 where $|f\ra_Q = \frac{1}{\sqrt2} (|\psi_1\ra_Q  - i |\psi_2\ra_Q)$ corresponds to a fringe pattern and $|a\ra_Q = \frac{1}{\sqrt2} (|\psi_1\ra_Q + i |\psi_2\ra_Q)$ corresponds to an antifringe (phased-shifted) pattern. Now, $\rho^0_Q = |f\ra_Q\la f|$ and $\rho^1_Q = |a\ra_Q\la a|$ are coherent superpositions, i.e., it is possible to extract interference on the screen. At this angle, $S(Q : D_B) = 1$. Therefore, the mutual entropy $S(Q:D_B)$ is related to the coherence of the conditional states $\rho^k_Q$, and in turn, to the visibility of interference fringes as we will see below.


 {\em Path information.---}From the joint density matrix $\rho_{Q D_A}$ computed in Eq.~\eqref{rhoAsDA}, the polarization measurement with detector $D_A$ reveals nothing about the spatial degree of freedom of $A$ since the joint state is completely decoupled. It follows that the mutual information vanishes 
 \beq
    S(Q : D_A) = 0.
 \eeq 
 In other words, if we do not know the outcome of the polarization measurement of photon $B$, an attempt to measure the polarization of $A$ after it traverses the double slit and QWPs will not reveal anything about the spatial state of $A$. On the other hand, if we do know the state of $D_B$, then the conditional mutual information is 
 \beq
    S(Q : D_A | D_B) = S \ge 0, \label{whichpath}
 \eeq 
 and varies with the erasure angle $\theta$.

 To understand the behavior of this quantity as a function of $\theta$, consider the state of $Q$ and $D_A$, conditional on the outcome $k$ of $D_B$. According to Eq.~\eqref{As-DADB}, this state is $\rho^k_{Q D_A} = \frac12 \sum_m |\psi^k_m\ra_Q\la\psi^k_m| \otimes |m\ra_{\!D_A\!}\la m|$. For an outcome $k=0$ of a polarization measurement of photon $B$ at angle $\theta = 0$,
 \beq
      \theta\!=\!0\!: \rho^0_{Q \!D_{\!A}} \!\!=\! \frac12 \Big[ |\psi_1\ra_{\!Q\!}\la\psi_1| \otimes |L\ra_{\!D_{\!A\!}\!}\la L| + |\psi_2\ra_{\!Q\!}\la\psi_2| \otimes |R\ra_{\!D_{\!A\!}\!}\la R|\Big]\!.
 \eeq
 At this angle, the conditional mutual information is maximal, $S(Q : D_A | D_B) = 1$, and it is clear that the state of the polarization detector $D_A$ is associated with one of the states $|\psi_j\ra_Q$, so that the measurement can reveal information about the path of photon $A$. For instance, an outcome $L$ corresponds to the state $|\psi_1\ra_Q$. As the erasure angle $\theta$ increases from zero to $\pi/4$, the information we have about the spatial state of $A$ is reduced to zero, since the density matrix becomes decoupled:
 \beq
      \theta = \frac{\pi}{4}\!: ~~\rho^0_{Q D_{\!A}} \!=\! |f\ra_Q\la f| \otimes \frac12  \mathbbm{1}_{D_A}.
 \eeq
 At this angle, $S(Q : D_A | D_B) = 0$. Therefore, the tagging operation with the QWPs only reveals information about the path of $A$ as long as we have additional information about the state of photon $B$ from its polarization measurement with detector $D_B$. Thus, Eq.~(\ref{whichpath}) is the correct expression for which-path information. Note further that
 \beq
    S(Q:D_AD_B) = 1\;,
 \eeq 
 which implies that---given the outcomes of both polarization detectors $D_A$ and $D_B$---it is possible to predict 
 {\em with certainty} the outcome of a direct measurement of the path of photon $A$.

 \section{Discussion}
 The quantities considered in the previous section can be used to generalize the usual concepts of coherence and path information, allowing us to construct a more fundamental relationship that is derived from information-theoretic principles and a no-collapse model of quantum measurement~\cite{CerfAdami1998}. 
 
 Recapitulating the results from the previous section, we know that whether we extract fringes or antifringes from the screen is controlled by the state of $D_B$. The visibility of the fringes is, in turn, related to the coherence of the conditional state $\rho^k_Q$ of $Q$ in Eq.~\eqref{rhoAs-k} and the mutual information $S(Q:D_B)$. As we have already seen, at angle $\theta = 0$ ($\theta = \pi/4$) the two variables $Q$ and $D_B$ are completely uncorrelated (correlated), the conditional state $\rho^k_Q$ is a statistical mixture (coherent superposition), and we observe no interference (full interference).

 On the other hand, information about the path of photon $A$ is determined by the correlation between its polarization (via the state of detector $D_A$) and its spatial states. This correlation is computed from the conditional mutual information $S(Q \!:\! D_A | D_B)$, and must be conditioned on the state of $D_B$ since $Q$ and $D_A$ are otherwise uncorrelated. When $\theta = 0$ ($\theta = \pi/4$), the variables $Q$ and $D_A$ are completely correlated (uncorrelated) given $D_B$, so that $D_A$ can (cannot) reveal path information, and we extract no interference (full interference) from the screen.

    \begin{figure}[t]
      \centering
      \includegraphics[width=0.78\linewidth]{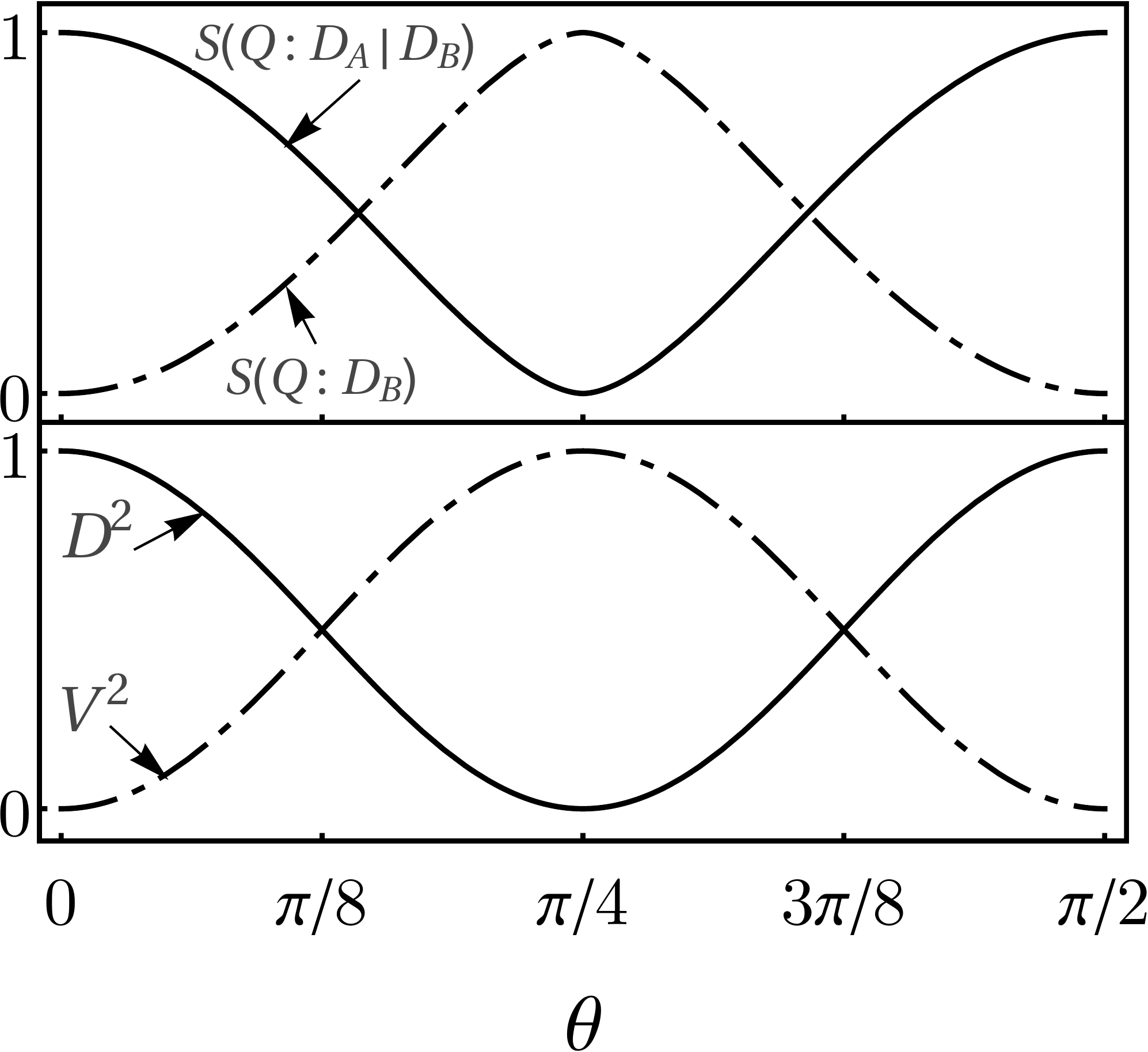}~~
      \caption{Top: Relationship between coherence and path information in Eq.~\eqref{info} for the Bell-state quantum eraser as a function of the erasure angle $\theta$. Shown are the information-theoretic quantities for the path information $S(Q\!:\!D_A|D_B) = S$ (solid) and coherence $S(Q\!:\!D_B) = 1 - S$ (dashed). Bottom: The distinguishability $D$ (solid) and fringe visibility $V$ (dashed), defined below, as a function of the erasure angle $\theta
      $. When $\theta = 0$ ($\theta = \pi/4$) there is full (no) path information and no (full) coherence.}  \label{fig:tradeoff}
    \end{figure}


 These two information-theoretic quantities, namely, the quanton coherence $S(Q \!:\! D_B)$ and the path information $S(Q \!:\! D_A | D_B)$, are fundamentally linked through the chain rule for entropies
 \beq\label{info}
    S(Q \!:\! D_B) + S(Q \!:\! D_A | D_B) = S(Q \!:\! D_A D_B) = 1,
 \eeq 
 and their sum is conserved throughout the erasure process. This information-theoretic formulation of complementarity generalizes earlier attempts~\cite{WoottersZurek1979,AngeloRibeiro2015,Baganetal2016} by explicitly referencing the measurement devices.  We note that the absence of correlations between detectors $D_A$ and $D_B$, $S(D_A:D_B)=0$, is crucial to enforce complementarity.

 We show in the top of Fig.~\ref{fig:tradeoff} the coherence and information in Eq.~(\ref{info}) as a function of the erasure angle $\theta$. We can compare these expressions to two other measures that are commonly used to discuss the wave-particle duality, namely the  distinguishability $D$ and the visibility $V$~\cite{greenberger1988,englert1996}. In general, $D^2 + V^2 \le 1$, but becomes an equality when the detectors are prepared in pure states. In the particular case we are discussing, $D^2 \!=\! \cos^2(2\theta)$, while the fringe visibility $V^2 \!=\! \sin^2(2\theta)$. They are shown in the bottom part of Fig.~\ref{fig:tradeoff} and exhibit a remarkably similar behavior when compared to the information-theoretic complementarity principle. %

 From Eq.~\eqref{info}, we can derive additional information-theoretic relations with conditional and mutual entropies. With the definition of conditional mutual information~\cite{CerfAdami1998}, $S(Q \!:\! D_A | D_B) = S(Q | D_B) - S(Q | D_A D_B) = S(Q | D_B)$, Eq.~\eqref{info} becomes a relation between a mutual information and a conditional entropy:
 \beq\label{info2}
    S(Q \!:\! D_B) + S(Q | D_B) = 1.
 \eeq 
 Furthermore, $S(Q\!:\!D_A) - S(Q\!:\!D_A|D_B) = S(Q\!:\!D_A\!:\!D_B) \le 0$, so that Eq.~\eqref{info} becomes
 \beq\label{inequality-1}
    0 \le  S(Q \!:\! D_A) + S(Q \!:\! D_B) \le 1,
 \eeq 
 where the the lower bound comes from the non-negativity of mutual entropies. This can be rewritten in terms of conditional entropies as
 \beq\label{inequality-2}
   1 \le  S(Q | D_A) + S(Q | D_B) \le 2,
 \eeq 
 where we used $S(Q) = 1$.


 Bagan et al. recently constructed an entropic complementarity relation between coherence and path information in an interferometer~\cite{Baganetal2016} using an entropic measure for coherence. For path states with equal probability, orthogonal detector states, and orthogonal measurements, their relation can be written as an equality very similar to ours
 \beq \label{hillery}
        C_{{\rm rel \, ent}}(\rho) + H(M:D) = 1 ,
 \eeq 
 where $C_{{\rm rel \, ent}}(\rho) = 1 - S(\rho)$ is a measure of the coherence of the particle's state $\rho$ in the interferometer, and $H(M:D) = S(\rho)$ is the path information, which is the mutual entropy of the path detector states $D$ and the results $M$ of probing them with a measurement. The connection to our result~\eqref{info} is immediately obvious, as $S(\rho)$ in Eq.~\eqref{hillery} is indeed equivalent to our $S$, the entropy of the conditional state $\rho^k_Q$ of photon $A$ in Eq.~\eqref{rhoAs-k}. However, our measures of coherence and path information emerge naturally from a full information-theoretic analysis and yield more insight into the origins of their complementarity, in particular how the quanton's entropy is distributed among the detectors $D_A$ and $D_B$ (as summarized by Fig.~\ref{fig:venn}).

 \section{Conclusions}
 We prefer to tread lightly when using our results to discuss aspects of quantum theory that have been discussed in a controversial manner since the discussions between Bohr and Einstein concerning these matters~\cite{WheelerZurek1984}. Nevertheless, we believe some statements can be made unequivocally. For example, it is now clear (and has been pointed out repeatedly before us), that a quanton not only carries both particle and wave attributes, but that these quantities are manifested in measurement devices in a fluid manner. In particular, the dynamics of the Bell-state quantum eraser, which allows us to give measurements different ``meanings" depending on what state preparation we may choose {\em after} the state determination has taken place, cannot possibly be consistent with a picture of quantum measurement in which the quantum state is irreversibly projected so as to be consistent with the state of the measurement device. The actuality of not only information erasure, but the production of alternative outcomes via the retroactive manipulation of state preparation, confirms the picture that the quanton wavefunction after measurement continues to carry amplitudes that are {\em not} consistent with the state of the measurement device. 
 
 That the quantum state can be inconsistent with the state of the measurement device should not come as a surprise to practitioners of quantum information science. After all, the idea of the classical measurement, in which the state of the system to be measured is copied onto the state of the measurement device, cannot carry over to quantum mechanics on account of the no-cloning theorem~\cite{WoottersZurek1982,Dieks1982}. Indeed, the central idea of classical measurement---in which the variation of the system is fully correlated with the variation in the device---is impossible for pure quantum states that carry no entropy whatsoever.
 
 Of course, mixed quantum states (pure joint states with a reference state traced out) can carry entropy, and this entropy can be shared with classical measurement devices. This appears to be the case in the construction described here, as the entropy of the quanton is exactly one bit (in the ideal case whose extension is discussed in footnote [16]). 
 If the classical device (here the device $D_A$) cannot carry information about the quanton's state, what information does it reflect? In our view, a classical device's state must be consistent with the state of other classical measurement devices, so as to ensure a causally consistent world. Here, the information $S(Q:D_A|D_B)$ predicts the outcome of a measurement of the which-path information that would be obtained if a device was placed squarely in the path of the beam. Of course, such a device would record a random outcome (the photon would be found in state $\psi_1$ half the time), and $D_A$ would perfectly predict this random outcome. Still, neither of these states predict the quanton's state, which after all is neither here nor there. We are thus forced to admit that our classical devices do not (and {\em cannot}) reveal to us the quantum reality underlying our classical world~\cite{AdamiCerf1999}. However, experimental (and theoretical) ingenuity has allowed us to be aware of our classical device's deceptions, and shown us the path to perhaps design even more clever schemes to lift the veil from the underlying quantum reality.

 \begin{acknowledgments}
Financial support by a Michigan State University fellowship to JRG is gratefully acknowledged.
 \end{acknowledgments}

%
%

 \appendix*
 
 \section{Interference patterns}
 Here we derive the spatial probability distributions for photon $A$ that is incident on a screen $D_X$ by modeling the interaction as a von Neumann measurement of the spatial location of photon $A$. Expanding the spatial states of $A$ in terms of the position basis of the screen yields
 \beq
    |\psi_j\ra_Q = \sum_{x=1}^n \psi_j(x) \, |x\ra_Q,
 \eeq 
 where $j=0,1$ labels each slit. The states $|x\ra$ can be discretized into $n$ distinct locations according to 
 \begin{eqnarray}
    |x=1\ra &=& |100 \cdots 0\ra, \nonumber \\
    |x=2\ra &=& |010 \cdots 0\ra, \nonumber\\
            &\vdots & \nonumber \\
    |x=n\ra &=& |0 \cdots 001\ra, \nonumber
 \end{eqnarray}
 which denote the location $x$ at which a photon is detected by $D_X$. Inserting this basis into the expression~\eqref{AB-DB} and performing the measurement of $Q$ with $D_X$ (which starts in the initial state $|x=0\ra = |0 \cdots 0\ra$), we come to
 \beq\label{AB-DXDB}
    |\Psi'\ra_{\!AD_{X}\!BD_{B}\!\!} = \frac12 \! \sum_{xmk} \! i^m  \psi^k_m(x)  |xx\ra_{QD_{X}\!\!} \otimes |m\ra_{\!P\!} \otimes |kk\ra_{\!BD_{B}},
 \eeq
 where we define the amplitudes $\psi_m^k(x) = \la x |\psi_m^k\ra_Q$.


 Tracing over $A$ and $B$ in Eq.~\eqref{AB-DXDB}, we arrive at the (classical) joint density matrix for detector $D_B$ and the screen
 \beq\label{rhoDBDX}
 \begin{split}
    \rho_{D_XD_B} & = \frac14 \sum_{xmk} |\psi^k_m(x)|^2 \, |x\ra_{\!D_X\!}\la x| \otimes |k\ra_{\!D_B\!}\la k|\\
                  & = \frac12 \sum_{k} \rho^k_{D_{X}} \! \otimes  |k\ra_{\!D_{B}\!}\la k|,
 \end{split}
 \eeq
 where
 \beq \label{rhoDX-k}
    \rho^k_{D_X} \! = \sum_x p_k(x) \, |x\ra_{\!D_X\!}\la x| ,
 \eeq 
 are the conditional states of the screen $D_X$ with corresponding conditional probability distribution
 \beq\label{pk}
    p_k(x) = \frac12 \sum_m |\psi_m^k(x)|^2 .
 \eeq 
 Tracing out detector $D_B$ from~\eqref{rhoDBDX} yields the full density matrix for $D_X$
 \beq
    \rho_{D_X} = \frac12 \sum_k \rho^k_{D_{X}} =  \sum_x p(x) \, |x\ra_{\!D_{X}\!}\la x|,
 \eeq 
 where the total probability distribution of the screen is
 \beq
    p(x) = \frac12 \sum_k p_k(x).
 \eeq


 It is straightforward to show that the total probability distribution $p(x)$ for the screen is completely incoherent due to the cancellation of the cross terms of the two conditional probabilities. That is,
 \beq
    p(x) =  \frac12 \Big( |\psi_{1}(x)|^2 + |\psi_{2}(x)|^2 \Big).
 \eeq 
 This distribution describes two intensity peaks on the screen corresponding to the pattern obtained from each slit individually. From the data as a whole (i.e., when we do not know the outcome of detector $D_B$) no interference is observed on the screen.

 However, suppose we do know the outcome of the polarization measurement of photon $B$. For an outcome $k$, the conditional state of the screen $D_X$ is given by Eq.~\eqref{rhoDX-k}. To discern the type of interference pattern the probability distribution~\eqref{pk} of this density matrix describes, we rewrite the conditional probability in terms of the original amplitudes $\psi_j(x) = \la x| \psi_j\ra_Q$, leading to
 \beq\label{interference}
 \begin{split} 
  p_k(x) & \!=
         \!\frac12 \bigg[|\psi_1(x)|^2 + |\psi_2(x)|^2 \!+  i \, (-)^k \sin2\theta \\
         & \qquad ~~~~~ \times \!\Big[\psi_1(x) \, \psi_2(x)^* \!- \psi_1(x)^* \, \psi_2(x)\Big]\bigg].
 \end{split} 
 \eeq
 where we used $U_{1k} \, U_{0k}^* + U_{1k}^* \, U_{0k} = (-)^k \sin2\theta$. In general, this expression will describe interference fringes with a visibility that is controlled by the magnitude of the coefficient $\sin2\theta$ in front of the cross terms. Let us consider two specific cases of the erasure angle $\theta$.

 First, $\theta=0$ corresponds to a measurement of photon $B$ in the linear $|h\ra,\,|v\ra$ basis. In this case, expression~\eqref{interference} reduces to an incoherent sum
 \beq
    \theta=0: ~~ p_k(x) = \frac12  \Big( |\psi_1(x)|^2 + |\psi_2(x)|^2 \Big),
 \eeq 
 and describes two intensity peaks on the screen with no interference. In turn, we have full information about the path of photon $A$. Second, $\theta=\pi/4$ describes a measurement of photon $B$ in the diagonal $|\pm\ra = \frac{1}{\sqrt2} (|h\ra \pm |v\ra)$ basis. In this case, expression~\eqref{interference} becomes a perfectly coherent sum
 \beq
    \theta = \frac{\pi}{4}: ~~ p_k(x) = \frac12  \left|\psi_1(x) - i\, (-)^k \psi_2(x)\right|^2 .
 \eeq 
 That is, the effect of tagging has been erased since the standard double-slit diffraction pattern can be observed. In general, given an outcome $k$ for detector $D_B$, the corresponding state of the screen is $\rho^{\,k}_{D_{\!X}}$ with probability distribution $p_k(x)$. This leads to fringes with a level of visibility that is determined by the erasure angle $\theta$. The distribution for $k=0$ is phase shifted relative to $k=1$, so that depending on the state of $D_B$, one observes either fringes or antifringes. Therefore, measuring photon $B$ in a basis characterized by the angle $\theta$ allows one to tune the visibility of the interference fringes from the standard two-slit diffraction to single-slit diffraction~\cite{kaiser2012}.

 \begin{figure}[t]
      \centering
      \includegraphics[width=0.56\linewidth]{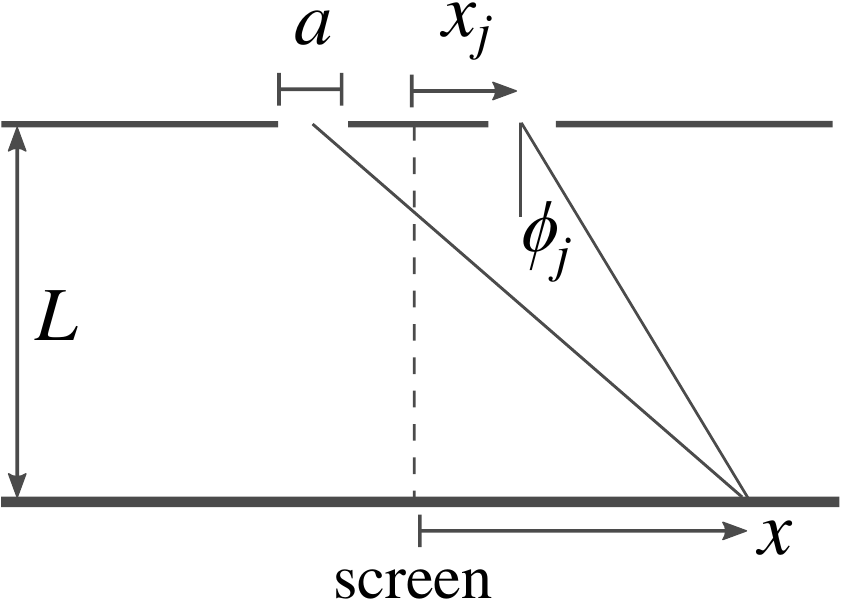}
      \caption{Geometry of the double slit apparatus. The slit width is $a$, the distance from the origin to the center of slit $j$ is $x_j$, the distance from the slits to the screen is $L$. The angle from the center of slit $j$ to point $x$ on the screen is $\tan\phi_j = (x-x_j)/L$.}
      \label{fig:geometry}
 \end{figure}

 \begin{figure}[t]
      \centering
      \includegraphics[width=0.825\linewidth]{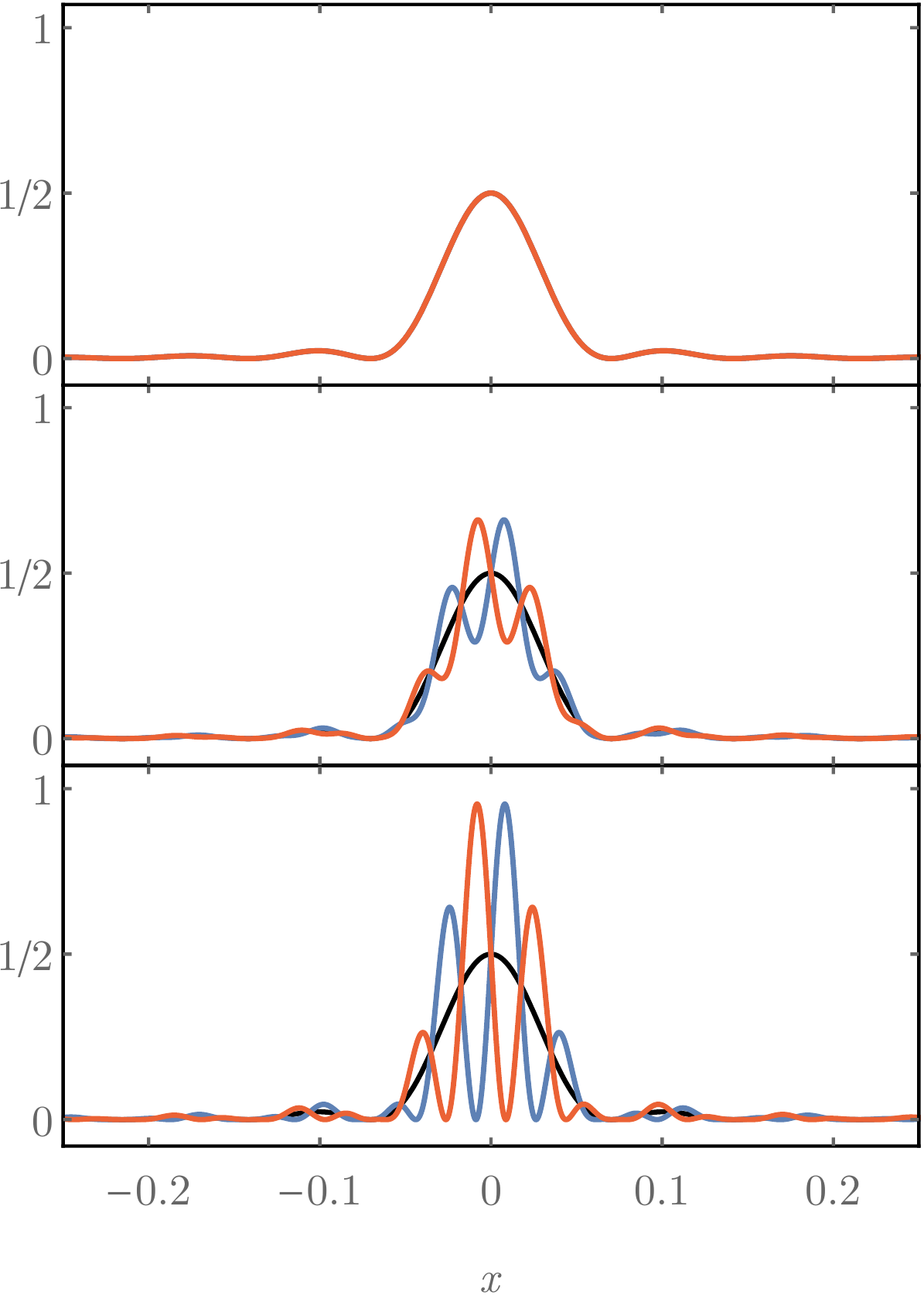}\hspace{0.2in}
      \caption{Conditional probability distributions $p_k(x)$, normalized so that the maximum is at 1. The blue (red) oscillations describe the interference pattern $p_0(x)$ ($p_1(x)$) of photon $A$ conditional on outcome $k=0$ ($k=1$) of detector $D_B$. The black line is the total distribution $p(x)$ and is the single-slit diffraction result. Parameters for this specific case are $a = 10 \,\mu$m, $d = 20 \,\mu$m, $L=1$ m, and $\lambda = 702$ nm. Top to bottom: probability distributions for three erasure angles $\theta = 0$, $\theta=\pi/16$, $\theta = \pi/4$.}
      \label{fig:interference}
 \end{figure}

 To explicitly compute the interference patterns, we write the amplitudes of the $j$th slit for a photon of wavelength $\lambda$ as~\cite{arsenovic2002}
 \beq
    \psi_j(x) \!=\! \sqrt\frac{a}{2\pi}  \,\, \frac{\sin\alpha}{\alpha} \,\, e^{- 2 i\,\alpha  \, x_j / a } \, ,
 \eeq
 where $\alpha = \pi a \sin\phi_j / \lambda$. The geometry of the double-slit apparatus (see Fig.~\ref{fig:geometry}) is described by the slit width $a$, the distance $x_j$ to the center of the $j$th slit, the angle $\phi_j = \tan^{-1}((x-x_j)/L)$ from the center of slit $j$ to the position $x$ on the screen, and the distance $L$ from the slits to the screen. For a single slit at the origin, the probability to detect a photon at position $x$ on the screen is
 \beq
    |\psi_j(x)|^2 = \frac{a}{2\pi}  \, \left| \frac{\sin\alpha}{\alpha} \right|^2 \, ,
 \eeq 
 which is the standard result for single-slit Fraunhofer diffraction. For two slits separated by a distance $d = x_2 - x_1$, the amplitudes for each slit are coherently added. In the far-field limit $L \gg d$, we can use the approximation $\phi_j = \phi = \tan^{-1}(x/L)$. Using the amplitudes $\psi_j(x)$ in the expression~\eqref{interference} for $p_k(x)$, leads to the interference patterns shown in Fig.~\ref{fig:interference}. The two patterns $p_0(x)$ and $p_1(x)$ are shifted relative to each other on the screen, and the envelope of each pattern is a single-slit diffraction pattern. We show the distributions for three erasure angles: $\theta = 0, \, \pi/16, \, \pi/4$. For a measurement in the linear $|h\ra$, $|v\ra$ basis ($\theta = 0$), there is no interference on the screen, since there is full information about the path of photon $A$. As the erasure angle increases to $\pi/4$ (a measurement in the diagonal $|\pm\ra$ basis), the oscillations increase to the level of the usual interference pattern for two-slit diffraction. The solid black line is the total distribution $p(x)$, which is the full data observed in the experiment, and shows no interference. Only by knowing the outcome $k$ of detector $D_B$ can one extract the associated conditional probability distribution $p_k(x)$ from the full distribution $p(x)$.

 \bibliography{eraser}  
 
\end{document}